\documentclass[11pt,a4paper]{iopart}
\usepackage{iopams}  
\usepackage{bm}
\begin{document}
\jl{14}
\newcommand{\be}{\begin{equation}}
\newcommand{\ee}{\end{equation}}
\newcommand{\ba}{\begin{eqnarray}}
\newcommand{\ea}{\end{eqnarray}}
\newcommand{\onehalf}{\textstyle{\frac{1}{2}}}

\title[]{Lorentz transformations in de Sitter relativity}

\author{R Aldrovandi, J P Beltr\'an Almeida, C S O Mayor and J G Pereira}

\address{Instituto de F\'{\i}sica Te\'orica, 
Universidade Estadual Paulista, Rua Pamplona 145, 01405-900 S\~ao 
Paulo, Brazil}

\begin{abstract}
The properties of Lorentz transformations in de Sitter relativity are studied. It is shown that, in addition to leaving invariant the velocity of light, they also leave invariant the length--scale related to the curvature of the de Sitter spacetime. The basic conclusion is that it is possible to have an invariant length parameter without breaking the Lorentz symmetry. This result may have important implications for the study of quantum kinematics, and in particular for quantum gravity.
\end{abstract}

\section{Introduction}

Our understanding of elementary particle physics is intimately related to both group representations and special relativity. In fact, all particles of Nature can be classified according to the irreducible representations of the Poincar\'e group ${\mathcal P}$, the kinematic group of special relativity. This property suggests that the symmetry of special relativity must be considered as an exact symmetry of Nature, a fact corroborated by many experiments. In principle, therefore, there is no reason to replace Poincar\'e as the kinematic group of spacetime. However, when one tries to merge elementary particle physics --- or quantum field theory --- with gravitation, one faces conceptual problems related to the existence of a limit length--scale, given by the Planck length. This scale, as is well known, shows up as the threshold of a new physics, represented by quantum gravity. Now, whether gravity is classical or quantum cannot depend on the observer. This means that the Planck length --- or some fundamental scale related to it --- must remain invariant under the relevant kinematics ruling the high--energy physics near the Planck scale. Since in ordinary special relativity a length will contract under a Lorentz boost, the invariance requirement of such length scale seems to indicate that either the Lorentz symmetry must be broken down, or the action of the Lorentz group ${\mathcal L}$ --- in particular the boosts --- must someway be modified~\cite{maj-smo-ame}.\footnote{To comply with this hypothesis, several deformed special relativities have been proposed, as for example those based on a $\kappa$-deformed Poincar\'e group~(see Ref.~\cite{dsr} and references therein), or those based on linear realizations of generalized groups~\cite{das}.} That is to say, near the Planck scale, the Poincar\'e group must be replaced by a more general group, which will preside over the high--energy kinematics. 

On the other hand, in the presence of a cosmological constant $\Lambda$, Minkowski spacetime $M$ is no longer a vacuum solution of the $\Lambda$--modified Einstein equation. If we assume that $\Lambda$ has a purely geometric nature, and that Einstein equation defines all acceptable spacetimes, Minkowski space becomes physically meaningless. In fact, for a positive $\Lambda$, absence of gravitation turns out to be represented by the de Sitter spacetime~\cite{ellis} --- here denoted $dS(4,1)$ --- whose kinematics is ruled, not by the Poincar\'e group ${\mathcal P}$, but by the de Sitter group $SO(4,1)$. In this case, ordinary Poincar\'e special relativity will no longer be valid, and must be replaced by a de Sitter special relativity~\cite{aap}.\footnote{Similar ideas have been explored in reference~\cite{getall}.}

Like Minkowski, defined as the quotient space between Poincar\'e and the Lorentz groups,
\[
M = {\mathcal P}/{\mathcal L},
\]
the de Sitter spacetime is also a homogeneous space~\cite{livro}:
\[
dS(4,1) = SO(4,1) / {\mathcal L}.
\]
Similarly to ordinary special relativity, therefore, in a de Sitter special relativity the Lorentz subgroup remains responsible for both the isotropy of space (rotation group) and the equivalence of inertial frames (boosts). The four additional transformations, given by a combination of translations and proper conformal transformations~\cite{abap}, define the homogeneity of de Sitter spacetime.

Given the above considerations, the purpose of this paper will be to explore the relation between equivalent frames in the context of a de Sitter relativity~\cite{aap}. In other words, we are going to study the properties of Lorentz transformations taking place in a de Sitter spacetime. We begin by reviewing, in the next section, the basic features of de Sitter transformations. 

\section{The de Sitter transformations}

The de Sitter spacetime $dS(4,1)$ can be viewed as a hyper-surface in the pseudo-Euclidean ambient space ${\bf E}^{4,1}$, an inclusion whose points in Cartesian coordinates $(\chi^A) = (\chi^0, \chi^1, \chi^2, \chi^3, \chi^{4})$ satisfy
\be
\eta_{AB} \chi^A \chi^B \equiv (\chi^0)^2 - (\chi^1)^2 -
(\chi^2)^2 - (\chi^3)^2 - (\chi^{4})^2 = -\, l^2,
\label{dspace1}
\ee
with $l$ a length parameter related to the spacetime curvature. In terms of these coordinates, an infinitesimal de Sitter transformation is written as
\be
\delta \chi^A = -\,\, {\mathcal E}^{A}{}_B \, \chi^B,
\ee
with ${\mathcal E}^{A}{}_B$ the transformation parameters. By considering a stereographic projection from the de Sitter hyper-surface into a target Minkowski spacetime, the above transformation can be written in a four-dimensional form. Using the Latin alphabet ($a, b, c \dots = 0,1,2,3$) to denote the four-dimensional algebra and Minkowski space indices, relation (\ref{dspace1}) can be rewritten as
\be
\eta_{a b} \, \chi^{a} \chi^{b} - (\chi^4)^2 = - \, l^2,
\label{dspace2}
\ee
where $\eta_{a b} = $ diag $(1$, $-1$, $-1$, $-1)$. Denoting then the Minkowski--valued stereographic coordinates by $x^a$, such a projection is defined by~\cite{gursey}
\be
\chi^{a} = \Omega(x) \, x^a \quad \mbox{and} \quad \chi^4 = -\, l \, \Omega(x) \left(1 + \frac{\sigma^2}{4 l^2} \right),
\label{stepro}
\ee 
where
\be
\Omega(x) = \frac{1}{1 - {\sigma^2}/{4 l^2}},
\label{n}
\ee
with $\sigma^2 = \eta_{a b} \, x^a x^b$ a Lorentz--invariant squared length. When written in terms of the stereographic coordinates, the de Sitter line element is
\be
d\tau^2 = g_{ab} \, dx^a dx^b,
\label{dsle}
\ee
with
\be
g_{ab} = \Omega^2 \, \eta_{ab}
\label{dsim}
\ee
the Minkowski--valued de Sitter metric. In stereographic coordinates, therefore, the de Sitter metric assumes a conformally flat form.

In terms of the stereographic coordinates, a de Sitter transformation appears on the target Minkowski spacetime as
\be
\delta x^a = - \, \epsilon^a{}_b \, x^b - \epsilon^b \Big[\delta^a_b -
\frac{1}{4 l^2} \left( 2 \, \eta_{bc} \, x^a x^c - \sigma^2 \, \delta^a_b \right) \Big],
\label{dstrans1}
\ee
where we have used the identifications~\cite{gursey}
\be
\epsilon^{ab} = {\mathcal E}^{ab} \quad \mbox{and} \quad
\epsilon^{a} = l \, {\mathcal E}^{a4}.
\ee
It can also be written in the form
\be\label{ds-t}
\delta x^a = \onehalf \epsilon^{cd} L_{cd} x^a - \epsilon^b \Pi_b x^a,
\ee
where
\be\label{lorentz}
L_{ab} =
\eta_{ac} \, x^c \, P_b - \eta_{bc} \, x^c \, P_a
\label{dslore}
\ee
are the Lorentz generators, and
\be
\Pi_b \equiv \frac{L_{b 4}}{l} =
P_b - \frac{1}{4 l^2} \, K_b
\label{l0}
\ee
are the ``de Sitter translation'' generators, with
\be
P_a = {\partial}/{\partial x^a} \quad \mbox{and} \quad
K_a = \left(2 \eta_{ab} \, x^b x^c - \sigma^2 \delta_{a}^{c} \right) P_c,
\ee
respectively the translation and the {\it proper} conformal generators. We see from these expressions that, similarly to the Poincar\'e case, the generators of the de Sitter transformations can be decomposed into Lorentz generators plus four additional translation--like generators. The difference is that, whereas in the Poincar\'e group these additional generators correspond to ordinary translations, in the de Sitter case they are given by a combination of ordinary translations and proper conformal transformations~\cite{abap}. The breakdown of ordinary translation symmetry is the basic de\-vi\-ation of de Sitter relativity from ordinary special relativity~\cite{aap}.

\section{Strong equivalence principle}

In one of its versions, the strong equivalence principle states that, in the presence of a gravitational field, it is always possible to find a local coordinate system in which the laws of physics reduce to those of special relativity. When Poincar\'e special relativity is in mind, this is easily represented by noticing that at each point $p$ of a generic spacetime there is a tangent Minkowski space, which approximates the spacetime manifold in a small enough domain around $p$. On the other hand, as we have discussed, in the presence of a cosmological constant, ordinary special relativity is no longer valid, and must be replaced by a de Sitter special relativity. The local symmetry group of spacetime, therefore, changes from Poincar\'e to de Sitter, and the strong equivalence principle must change accordingly. Its modified version states that, in the presence of a gravitational field, it is always possible to find a local coordinate system in which the laws of physics reduce to those of a de Sitter special relativity.

Coordinates have always values in some open subset of an Euclidean (or pseudo--Euclidean) space, which is Minkowski space for the stereographic coordinates we are using. Equation~(\ref{dstrans1}) represents a de Sitter transformation on that Minkowski coordinate space. This is a mathematically convenient way to represent the local symmetry of spacetime also because, locally, one can always identify spacetime with the tangent Minkowski space. On each tangent Minkowski space, the laws of physics must be invariant under the de Sitter transformation (\ref{dstrans1}). It should be remarked, however, that this is not the only way to describe it. In fact, instead of tangent Minkowski spacetimes, one can consider osculating de Sitter spaces at each point $p$ of the spacetime manifold, whose isometries represent the local symmetry of spacetime. This is a different way to represent the local symmetry of spacetime. Since, in the presence of $\Lambda$, spacetime itself is a de Sitter space in the absence of matter, this second way of describing the local spacetime symmetry can be considered to be more convenient from the physical point of view. For our purposes here, it will be enough to consider the Minkowski realization of the de Sitter transformations.

\section{Lorentz transformation in de Sitter spacetime}

The Lorentz generators (\ref{dslore}) constitute the {\it orbital}\, part of the generators, whose complete form is
\be
J_{ab} = L_{ab} + S_{ab},
\label{eq:Jab}
\ee
with $S_{ab}$ the matrix {\it spin} part of the generators. For the specific case of a Lorentz vector, this matrix has entries~\cite{ramond}
\be
\left(S_{ab}\right)^c{}_d = \eta_{ad} \, \delta_b^c - \eta_{bd} \, \delta_a^c.
\ee
Generators $L_{ab}$ and $S_{ab}$ act on distinct spaces and (\ref{eq:Jab}) is actually a direct sum  of type $J = L \otimes I  \oplus S$, with $I$ the identity matrix. It is important to remark that, even when acting on de Sitter spacetime, these gen\-er\-ators still present the usual Lie algebra of the Lorentz group~\cite{aap}. This is a fundamental property in the sense that it allows the construction, on the de Sitter spacetime, of an algebraically well--defined special relativity. This possibility is related to the fact that, like Minkowski, de Sitter spacetime is homogeneous and isotropic~\cite{jack}.

An element of the Lorentz group is obtained by taking the exponential of the generators. For the specific case of the transformation of $x^a$, the appropriate group element can be written either in terms of the orbital generators $L_{ab}$ or in terms of the matrix generators $S_{ab}$. In terms of $L_{ab}$, it is
\be
\Lambda^c{}_d = \left\{\exp \left[{\textstyle \frac{1}{2}} \epsilon^{ab} \,  L_{ab} \right] \right\}\otimes \delta^c_d.
\ee
The corresponding finite Lorentz transformations has the form 
\be
x'^c = \Lambda^c{}_d \, x^d.
\label{LorenMink}
\ee
Now, the Lorentz group is defined as the group that leaves invariant the Minkowski metric:
\be
\eta_{ab} = \Lambda_a{}^c \, \Lambda_b{}^d \, \eta_{cd}.
\ee
Since the conformal factor $\Omega$ is Lorentz invariant, a Lorentz transformation leaves invariant also the de Sitter metric:
\be
g_{ab} = \Lambda_a{}^c \, \Lambda_b{}^d \, g_{cd}.
\ee
In a de Sitter spacetime, therefore, a Lorentz transformation leaves invariant both the speed of light $c$ and the length parameter $l$. It is important to remark that, differently from the $c$ invariance, which is essentially kinematic, the $l$ invariance has a geometric character.

As is well known, a Lorentz rotation in the plane $x t$ of Minkowski spacetime leaves invariant the difference
\be
c^2 t^2 - x^2,
\ee
with $t, x, y, z$ denoting the Cartesian coordinates. Using the same letters $t, x, y, z$ to denote the stereographic coordinates of the de Sitter spacetime, a Lorentz rotation in the plane $x t$ of this spacetime is found to formally coincide with that of ordinary special relativity:
\be
x' = \frac{x-vt}{\sqrt{1-v^2/c^2}}; \quad t' =\frac{t-vx/c^2}{\sqrt{1-v^2/c^2}}; \quad
y'=y; \quad z'=z.
\label{boost}
\ee 
An important feature of the stereographic coordinates is that they provide a linear realization of the Lorentz boosts both in coordinate and in momentum spaces. It is then easy to see how, in this simple modification of special relativity, the existence of an invariant length--scale does not require the breaking of the Lorentz symmetry.

\section{Final remarks}

A Lorentz transformation in de Sitter spacetime leaves invariant both the velocity of light $c$ and the de Sitter length parameter $l$. Since $l = \sqrt{\Lambda/3}$, a Lorentz transformation is found to leave invariant the cosmological constant $\Lambda$. In order to preserve a length scale, therefore, the Lorentz sym\-metry does not necessarily need to be broken down or modified. To a certain extent, this property could be considered trivial because we know since long from ordinary special relativity that a Lorentz transformation leaves invariant the vanishing cosmological constant associated to Minkowski spacetime. Although seemingly trivial, however, it can have important consequences for high--energy physics, and in particular for quantum gravity.
 
For example, if the phase transitions associated to the spontaneously broken symmetries are considered as the primary source for a non-vanishing $\Lambda$~\cite{carroll}, it turns out conceivable to assume that a high--energy phenomenon could modify the {\it local structure of space-time for a short period of time}, in such a way that the immediate neighborhood of a high energy phenomenon would depart from Minkowski and become a de Sitter --- or anti-de Sitter --- spacetime~\cite{mansouri}. There would then exist a connection between the energy scale of a given phenomenon and the local value of $\Lambda$. For a very high--energy phenomenon, the de Sitter length parameter $l$ would approach the Planck length, and the relevant local kinematics would be that of de Sitter relativity. According to this kinematics, different observers would see the same length parameter $l$.\footnote{For a hypothetical experiment with energy of the order of the Planck energy, the local value of $\Lambda$ would be of the order $\Lambda \sim 10^{66}~\mbox{cm}^{-2}$, which differs roughly from the observed cosmological constant by 120 orders of magnitude.}

This scenario fits quite reasonably with the idea that high energies might cause small--scales fluctuations in the texture of spacetime. The important point here is that the de Sitter relativity, together with the hypotheses of a connection between energy and the local value of $\Lambda$, {\it give a precise meaning to these local spacetime fluctuations}. Furthermore, due to the presence of a horizon at distance $l$, the local causal structure of spacetime, defined by the light cone, will also be modified at high energies in a very precise way~\cite{abap}. A de Sitter relativity, therefore, gives rise to a very precise high--energy phenomenology, opening up the door for an experimental confrontation. For example, it is known that local spacetime fluctuations can act as subatomic lenses, slowing down the propagation of very high--energy photons. In fact, recent experiments observed a delay in very high--energy gamma rays coming from extragalactic flare, in relation to the lower--energy ones~\cite{magic}. These data could eventually be used to probe the predictions of de Sitter special relativity, as well as of the hypotheses of a connection between energies and the local value of $\Lambda$.

\ack
The authors would like to thank FAPESP, CAPES and CNPq for partial financial support.


\end{document}